\documentclass[12pt]{article}
\usepackage{epsfig}

\def\beq{\begin{equation}}
\def\eeq#1{\label{#1}\end{equation}}
\def\eeqn{\end{equation}}

\def\beqa{\begin{eqnarray}}
\def\eeqa#1{\label{#1}\end{eqnarray}}
\def\eeqan{\end{eqnarray}}

\let\bar=\overbar

\def\Dslash{\not{\hbox{\kern-4pt $D$}}}
\def\dslash{\not{\hbox{\kern-2pt $\del$}}}

\def\msb{{\bar{\ssstyle M \kern -1pt S}}}

\def\Title#1{\begin{center} {\Large {\bf #1} } \end{center}}

\begin{document}

\Title{Diquark Properties and the TOV Equations
}

\bigskip\bigskip


\begin{raggedright}
{\it David Blaschke$^{a,}$\footnote{Postal address:
D-18051 Rostock, Germany; E-mail: david@darss.mpg.uni-rostock.de},
\underline{Sverker Fredriksson}$^{b,}$\footnote{Postal address:
SE-97187 Lule\aa , Sweden; E-mail: sverker@mt.luth.se}
and Ahmet Mecit \"Ozta\c{s}$^{ b,c,}$\footnote{Postal address:
TR-06532 Ankara, Turkey; E-mail: oztas@hacettepe.edu.tr}
\index{Fredriksson, S.}\\
}

\vskip 0.5cm
$^a$
{\it Department of Physics,
University of Rostock}\\
$^b$
{\it Department of Physics,
Lule\aa \ University of Technology}\\
$^c$
{\it Department of Physics,
Hacettepe University}
\bigskip\bigskip
\end{raggedright}


\section{Introduction: Diquarks}

This is a status report of our work
on quark/diquark effects inside compact
astrophysical objects. It goes somewhat in excess
of the results shown at the workshop.

The word diquark is due to
Gell-Mann in 1964 \cite{Gell-Mann64}.
The idea of a two-quark correlation
has now spread to many areas of particle
physics, motivated by phenomenology,
lattice calculations, QCD or
instanton theory.
Now there are some $1200$ papers on diquarks,
some of which were covered by a review in 1993
\cite{anselmino93}. The speaker (S.F.) has an
updated database with diquark papers.

Although there is no consensus
about diquark details,
it seems certain that a $(ud)$ quark pair
experiences some attraction when in
total spin-$0$ and colour-{\bf 3*}.
Nuclear matter should also
be subject to such pairing,
maybe in some new way,
{\it e.g.}, as in the nuclear
EMC effect \cite{fredriksson84}.
Quark pairing should also
affect a (hypothetical)
quark-gluon plasma (QGP),
and by now there are more than $300$ papers
built on this idea. Many of these
assume that the correlation
is like the one between electrons in a
superconductor, with diquarks being
the Cooper pairs of QCD
\cite{fredriksson83}. The notion of
``superconducting quark matter''
is due to Barrois in 1977 \cite{barrois77}.
Current efforts owe much to the
review paper by Bailin and Love
\cite{bailin84}, and to work by Shuryak, Wilczek and
collaborators \cite{rapp98,alford99}.
Diquarks in a QGP have also been analysed
``classically", with thermodynamics,
or field theory \cite{ekelin86,donoghue}.

{\it Astrophysical} diquarks gained popularity
about a decade ago, when they were
suggested to influence the
supernova collapse and ``bounce-off''
\cite{fredriksson89,kastor91,horvath,sahu93},
and to enhance the neutrino cooling
of quark-stars. The latter effect
is now subject to much
research \cite{voskresensky00,carter00}.

Here we study some features of compact
objects that would be sensitive to
diquark condensation in a QGP.
The form factor of the diquark correlation and the quark
isospin (a)symmetry due to presence of electrons will be
given special attention.

There are many situations
where a QGP with diquarks might play a
role: (i) In a {\it quark star}, which might
appear as dark matter \cite{witten84,fredriksson99};
(ii) In a {\it hybrid/neutron star},
surrounded by a hadronic crust;
(iii) In a {\it supernova}, or a
{\it `hypernova' gamma-ray burster},
where diquarks might trigger neutrino bursts
and the bounce-off;
(iv) In the {\it primordial plasma}
at the Big Bang, where diquarks might
have delayed the hadronisation.

\section{Formalism and Results}

We use the BCS theory of colour superconductivity
\cite{bailin84,rapp98,alford99}.
The gap $\Delta$
can be seen as the gain in energy due to the
diquark correlation.
Another gap, $\phi$, is related to the
quark-antiquark condensate.
The thermodynamical grand canonical
potential, $\Omega$, is minimised in its
variables, resulting in an equation of
state (EOS) and other relations.
We follow the approach of \cite{berges99}
for the $\Omega$ and generalise it for
isospin asymmetry between $u$ and $d$ quarks \cite{kiriyama01}
\begin{eqnarray}
\label{omega}
&&\Omega(\phi,\Delta;\mu_{B},\mu_{I},\mu_{e};T)= \nonumber \\
&&= \frac{\phi^{2}}{4G_{1}} + \frac{\Delta^{2}}{4G_{2}}
- \frac{1}{12\pi^{2}}\mu_{e}^{4}
-\frac{1}{6}\mu_e^{2}T^{2}-\frac{7}{180}\pi^{2}T^{4}\nonumber \\
&&-2\int_0^{\infty}\frac{q^{2}dq}{2\pi^{2}}(N_{c} - 2)\times
\left\{ 2E_{\phi}+\frac{}{}
    \right. \nonumber \\
&&+\left. T\ln \left[1 +\exp \left(-\frac{E_{\phi}- \mu_{B} -
\mu_{I}}{T}
\right) \right]
+T\ln \left[1+\exp \left(-\frac{E_{\phi} - \mu_{B}+ \mu_{I}}{T}
\right) \right] \right. \nonumber \\
&&\left. +T\ln \left[1 + \exp \left(-\frac{E_{\phi} + \mu_{B}-
\mu_{I}}{T}
\right) \right]
+T\ln \left[1 + \exp \left(-\frac{E_{\phi} + \mu_{B}+ \mu_{I}}{T}
\right) \right] \right\} \nonumber \\
&&-4\int_0^{\infty}\frac{q^{2}dq}{2\pi^{2}}\times
\left\{E_{-}+E_{+}+\frac{}{}
    \right. \nonumber \\
&&\left. +T\ln \left[1 +\exp \left(-\frac{E_{-}- \mu_{I}}{T} \right)
\right]
+T\ln \left[1 + \exp \left(-\frac{E_{-}+\mu_{I}}{T} \right) \right]
\right. \nonumber \\
&&+\left. T\ln \left[1 + \exp \left(-\frac{E_{+}-\mu_{I}}{T} \right)
\right]
+T\ln \left[1 + \exp \left(-\frac{E_{+}+\mu_{I}}{T} \right) \right]
\right\} + C~,
\end{eqnarray}
where the subtraction
$C = 
- \Omega((\phi_0^{vac})^{2},0;0,0,0;0)$
has been introduced to make (\ref{omega}) finite and to assure that
pressure and energy density of the vacuum at $T=\mu=0$ vanish.
Thus at the boundary of a compact quark matter object, where the
quark-condensate $\phi_{0}$ changes, a pressure difference arises,
which is necessary for confining the system, at least for small masses.

Instead of the chemical potentials $\mu_{u}$,
$\mu_{d}$ of $u$ and $d$
quarks, one uses \cite{kiriyama01}
those of baryon number, $\mu_{B} = (\mu_{u} + \mu_{d})/2$,  and
isospin asymmetry, $\mu_{I} = (\mu_{u} - \mu_{d})/2$.
Then, if beta equilibrium with electrons holds, $\mu_{e} = -2\mu_{I}$.
The particle densities are given by
$n_{B} = n_{u} + n_{d} = -{\partial \Omega}/{\partial \mu_{B}}$,
$n_{I} = n_{u} - n_{d} = -{\partial \Omega}/{\partial \mu_{I}}$
and
$n_{e} = - {\partial \Omega}/{\partial \mu_{e}}
= - {\mu_{e}^{3}}/({3\pi^{2}})$.
Charge neutrality means
$n_{B} + 3n_{I} - 6n_{e} = 0$,
and isospin symmetry $n_{I} = 0$.

The dispersion relations are
\begin{equation}
E_{\phi} = \sqrt{q^{2} + (m + F^{2}(q)\phi)^{2}}
\mbox{~~~~and~~~~}
E_{\pm} = (E_{\phi} \pm \mu_{B})
\sqrt{\displaystyle 1 + \frac{F^{4}(q)\Delta^{2}}
{(E_{\phi} \pm\mu_{B})^{2}}},
\end{equation}
\noindent where $F(q)$ is a form factor
for two-quark correlations.

We use $T = 0$ and $m = 0$,
and start by finding the gaps
$\phi_{0}$ and $\Delta_{0}$ that minimise
$\Omega$. These values give the EoS as
\begin{equation}
\Omega(\phi_{0},\Delta_{0};\mu_{B},\mu_{I},\mu_{e};T = 0)
= \epsilon -\mu_{B} n_{B} - \mu_{I} n_{I} - \mu_{e} n_{e} = -P,
\end{equation}
\noindent
where $\epsilon$ is the energy density
and $P$ the pressure.
\begin{figure}[t]
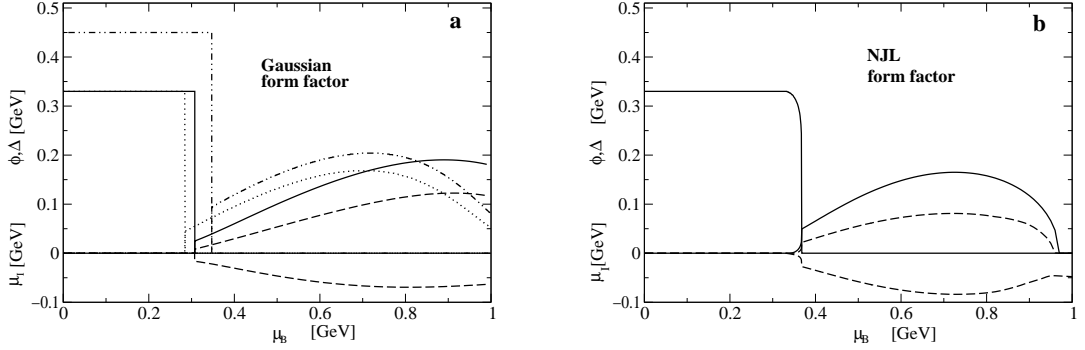

\begin{center}
\epsfig{file=Fig1aproc.eps,height=1.8in}
~~~~~~~~\epsfig{file=Fig1bproc.eps,height=1.8in}
\caption{\footnotesize{The gaps $\phi$, $\Delta$
and the isospin asymmetry chemical potential $\mu_{I}$
as functions of $\mu_{B}$ for (a) Gaussian, and (b)
NJL form factors.
{\it Solid lines} are with isospin symmetry
(and no electrons),
{\it Dashed lines} are with beta equilibrium.
{\it Dotted lines} in (a) are with isospin symmetry for
$\Lambda_{G}=0.8$ GeV and $\phi_{0}^{vac}=0.33$ GeV.
{\it Dashed double-dotted lines} in (a)
are with isospin symmetry for
$\Lambda_{G}=0.8$ GeV and $\phi_{0}^{vac}=0.45$ GeV
(giving the same vacuum pressure as with the solid lines).}}
\label{fig:1}
\end{center}
\end{figure}
We use the form factors
suggested by Schmidt {\it et al.} \cite{schmidt94},
who fitted those to masses and decay properties
of light mesons. The following special cases
are considered:
(i) Gaussian: $F^2(q)=\exp{(-q^2/\Lambda^2_G)}$;
(ii) Lorentzian: $F^2(q)= 1/[1+(q/\Lambda_L)^{4}]$;
(iii) cutoff NJL: $F^2(q)=\Theta(\displaystyle 1-q/\Lambda_{NJL})$.
Here $\phi_0=0.33$ GeV, $\Lambda_G=1.025$ GeV,
$\Lambda_L=0.8937$ GeV and $\Lambda_{NJL}=0.9$ GeV \cite{schmidt94}.
We contrast this to $\Lambda_G=0.8$ GeV used in \cite{berges99}.
Fig. 1 shows our results for the
Gaussian and NJL form factors,
and for two extreme cases:
(i) isospin symmetry (equal numbers of $u$ and $d$ quarks), and
(ii) beta equilibrium and charge neutrality.
In the former case charge
is balanced {\it outside} the object.
In the latter, the
electron fraction is small, and
$n_{d} \approx 2n_{u}$. Here $\mu_{I}$ is
given in the lower part of the figures.
We also test the sensitivity on
the parameter $\Lambda_{G}$.
There is now the choice to either keep
$\phi_{0}^{vac}$ fixed, or to change it so that the
vacuum pressure
${(\phi_{0}^{vac})^{2}}/{(4G_{1})}$
is kept fixed.

\newpage
The EoS is used as input
to the standard Tolman-Oppenheimer-Volkoff
equations \cite{oppenheimer39}
for an equilibrium (spherical)
fluid (see also \cite{glendenning00}):

\begin{equation}
\frac{dP(r)}{dr}=
-\frac{[\epsilon(r)+P(r)][m(r)+4\pi r^{3}P(r)]}{r[r-2m(r)]},
\end{equation}

\begin{equation}
m(r)=4\pi \int_{0}^{r}\epsilon(r')r'^{2}dr'.
\end{equation}

The equations are iterated in
$r$, starting with
some value $\epsilon_{0}$
at $r=0$. The procedure
stops when $P=0$, which defines the
radius $R$ of the object, and $M = m(R)$
is plotted {\it vs.} $R$, see Fig. 2.
Each $\epsilon_{0}$ value
results in one point in the plot, with unique values
also of $\mu_{B}$ and $\mu_{I}$. The graph
becomes a backbending spiral, with
$\epsilon_{0}$ and $\mu_{B}$ increasing along
the curve. Solutions on
left-going parts of the curve are
unstable.

\begin{figure}[t]
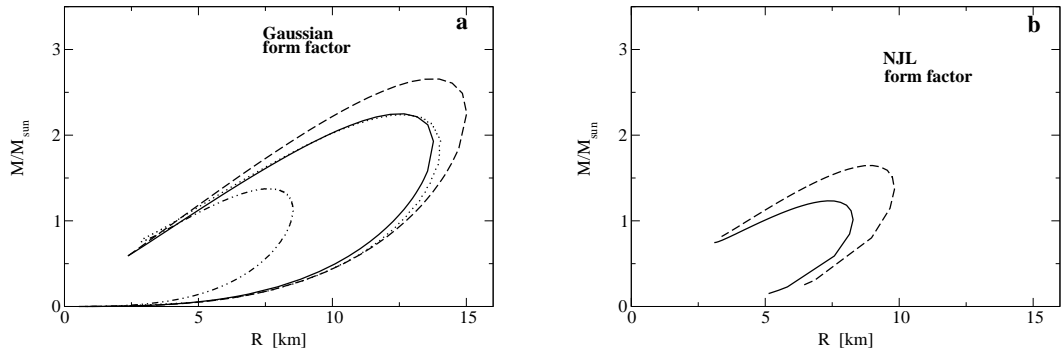
\label{fig:2}
\begin{center}
\epsfig{file=Fig2aproc.eps,height=1.8in}
~~~~~~~\epsfig{file=Fig2bproc.eps,height=1.8in}
\caption{\footnotesize{a) The mass ratio $M/M_{sun}$
{\it vs.} the radius $R$ of a compact object
according the the TOV equation.
The curves correspond to the same
situations as in Fig. 1.}}
\end{center}
\end{figure}

\section{Conclusions and Outlook}

The diquark form factor and the isospin
(a)symmetry are important inside
a quark/ diquark star. They can make
a difference of up to two solar
masses, as for its maximal mass.

We now have results also
for $0 < T < \Delta$,
where condensate-free states can
occur in intervals of $\mu_{B}$ \cite{ahmet01}.
These correspond
to an interaction-free plasma,
and appear as ``bumps'' in the TOV
curves. They might be relevant
as phase transitions inside
a collapsing system (supernova).
The QGP might suddenly enter a diquark phase,
whereafter it again
becomes a free-quark state just before the
bouncing \cite{hong01}.
We also intend to study such
events for the Big Bang plasma.
Maybe a transition
into a diquark phase resulted
in a release of neutrinos, just as
with the cosmic microwave background.
The neutrino energy would have
been of the order of $\Delta$,
but now cooled to keV energies.

\newpage

\noindent {\bf Acknowledgements}
\vskip0.1cm
\noindent SF and A\"O are grateful to the Organisers
for providing a very inspiring atmosphere. SF would like
to thank the University of Rostock,
and DB and A\"O the Lule\aa \ University
of Technology for hospitality during
visits. Some of these
have been supported by the European
Commission within the Erasmus programme.

\def\Discussion{
\setlength{\parskip}{0.3cm}\setlength{\parindent}{0.0cm}
         \bigskip\bigskip      {\Large {\bf Discussion}} \bigskip}
\def\speaker#1{{\bf #1:}\ }
\def\endDiscussion{}

\Discussion

\speaker{A. Thampan}
Should one not see a signature (say in terms
of phase transitions or so) in the $P$ {\it vs.} $\rho$
(EoS) relationships?

\speaker{Fredriksson}
For some extreme parameter values
bumps occur in, {\it e.g.},
$M$ {\it vs.} $R$ and $P$ {\it vs.} $\epsilon$,
and certainly also in a $P$ {\it vs.}
$\rho$, although we did plot the latter.
For the case of beta equilibrium,
these transitions are smoother in the TOV graphs.
The bumps are clearer for $T > 0$,
as mentioned above.

\speaker{J.E. Horvath}
Do diquarks disappear suddenly in this model? Could
you identify what makes the matter self-bound?

\speaker{Fredriksson}
Not suddenly in time, because these
are equilibrium solutions. But diquarks
disappear ``suddenly'' when other parameters change,
{\it e.g.}, $\mu_{B}$ (see ``NJL'').
A strange star is self-bound by gravity
at high masses, but also due to our
``vacuum pressure'' ($C$).

\speaker{Unknown}
Are the stars beyond the mass peak stable?

\speaker{Fredriksson}
Our approach does not tell, but states
with $\frac{\partial M}{\partial\epsilon} < 0$
are usually prescribed to be unstable \cite{glendenning00}.

\endDiscussion


\begin{thebibliography}{99}

\bibitem{Gell-Mann64}
M. Gell-Mann, Phys. Lett. {\bf 8}, 214 (1964).

\bibitem{anselmino93}
M. Anselmino {\it et al.}, Rev. Mod. Phys. {\bf 65}, 1199 (1993).

\bibitem{fredriksson84}
S. Fredriksson, Phys. Rev. Lett. {\bf 52}, 724 (1984).

\bibitem{fredriksson83}
S. Fredriksson, M. J\"andel and T.I. Larsson,
Phys. Rev. Lett. {\bf 81}, 2179 (1983).

\bibitem{barrois77}
B.C. Barrois, Nucl. Phys. {\bf B129}, 390 (1977).

\bibitem{bailin84}
D. Bailin and A. Love, Phys. Rep. {\bf 107}, 325 (1984).

\bibitem{rapp98}
R. Rapp {\it et al.},
Phys. Rev. Lett. {\bf 81}, 53 (1988).

\bibitem{alford99}
M. Alford, K. Rajagopal and F. Wilczek,
Phys. Lett. {\bf B450}, 325 (1999).

\bibitem{ekelin86}
S. Ekelin, in {\it Strong Interactions and Gauge Theories}, p. 559,
Ed. J. Tran Thanh Van (Editions Fronti\`eres, Gif-sur-Yvette, 1986).

\bibitem{donoghue}
J.F. Donoghue and K.S. Sateesh, Phys. Rev. {\bf D38}, 360 (1988);
K.S. Sateesh, Phys. Rev. {\bf D45}, 866 (1992).

\bibitem{fredriksson89}
S. Fredriksson, in {\it Workshop on Diquarks}, p. 22,
Eds. M. Anselmino and E. Predazzi (World Scientific,
Singapore, 1989).

\bibitem{kastor91}
D. Kastor and J. Traschen, Phys. Rev. {\bf D44}, 3791 (1991).

\bibitem{horvath}
J.E. Horvath, O.G. Benvenuto and H. Vucetich,
Phys. Rev. {\bf D44}, 3797 (1991);
O.G. Benvenuto, H. Vucetich and J.E. Horvath,
Nucl. Phys. Proc. Suppl. {\bf 24B}, 160 (1991);
J.E. Horvath, Phys. Lett. {\bf B294}, 412 (1992);
J.E. Horvath, J.A. de Freitas Pacheco and J.C.N. de Araujo,
Phys. Rev. {\bf D46}, 4754 (1992).

\bibitem{sahu93}
P. Sahu, Int. J. Mod. Phys. {\bf E2}, 647 (1993).

\bibitem{voskresensky00}
D. Blaschke, T. Kl\"ahn and D.N. Voskresensky,
Astrophys. J. {\bf 533}, 406 (2000);
D. Blaschke, H. Grigorian and D.N. Voskresensky,
A \& A {\bf 368}, 561 (2001).

\bibitem{carter00}
G.W. Carter and S. Reddy, Phys. Rev. {\bf D62}, 103002 (2000);
P. Jaikumar and M. Prakash,
Phys. Lett. {\bf B516}, 345 (2001).

\bibitem{witten84}
E. Witten, Phys. Rev. {\bf D30}, 272 (1984).

\bibitem{fredriksson99}
S. Fredriksson {\it et al.},
in {\it Dark Matter in Astrophysics and Particle Physics}, p. 651,
Eds. H.V. Klapdor-Kleingrothaus and L. Baudis
(Inst. of Physics, Bristol \& Philadelphia, 1998).

\bibitem{berges99}
J. Berges and K. Rajagopal, Nucl. Phys. {\bf B538}, 215 (1999).

\bibitem{kiriyama01}
O. Kiriyama, S. Yasui and H. Toki, hep-ph/0105170 (2001).

\bibitem{schmidt94}
S. Schmidt, D. Blaschke and Yu. Kalinovsky, Phys. Rev. {\bf C50}, 435 (1994).

\bibitem{oppenheimer39}
J.R. Oppenheimer and G. Volkoff, Phys. Rev. {\bf 55}, 377 (1939).

\bibitem{glendenning00}
N.K. Glendenning, {\it Compact Stars}
(Springer, New York \& London, 2000).

\bibitem{ahmet01}
D. Blaschke, S. Fredriksson, H. Grigorian and A. \"Oztas, in preparation.

\bibitem{hong01}
D.K.~Hong, S.D.~Hsu and F.~Sannino,
Phys. Lett. {\bf B516}, 362 (2001).

\end{thebibliography}
\end{document}